\documentclass{ws-ijgmmp}

\usepackage[breaklinks]{hyperref}
\usepackage{cite}
\usepackage[utf8]{inputenc}
\usepackage{color}
\usepackage{amsfonts}
\usepackage{bm,bbm}

\newcommand{\be}{\begin{equation}}
\newcommand{\ee}{\end{equation}}
\newcommand{\ba}{\begin{eqnarray}}
\newcommand{\ea}{\end{eqnarray}}
\def\bs{\begin{subequations}}
\def\es{\end{subequations}}
\renewcommand{\leq}{\leqslant}

\def\a{\alpha}

\def\g{\gamma}
\def\la{\lambda}
\def\k{\kappa}

\def\Om{\Omega}
\def\om{\omega}

\def\G{\Gamma}
  
\def\s{\sigma}

\def\vp{\varphi}

\def\cD{\mathcal{D}}

\def\cL{\mathcal{L}}

\def\cV{\mathcal{V}}

\def\ds{d_{\rm S}}
\def\dh{d_{\rm H}}
\def\dw{d_{\rm W}}

\def\p{\partial}

\newcommand{\Eq}[1]{(\ref{#1})}
\def\com{\color{magenta}}
\def\cob{\color{blue}}
\newcommand{\oarX}[1]{\href{http://arxiv.org/abs/#1}{{\ttfamily\com arXiv:#1}}}
\newcommand{\arX}[1]{\href{http://arxiv.org/abs/#1}{{\ttfamily\com arXiv:#1}}}
\newcommand{\doin}[6]{\href{http://dx.doi.org/#1}{{\cob {\it #2 #3} {\bf #4} (#6) #5}}}
\newcommand{\doinn}[5]{\href{http://dx.doi.org/#1}{{\cob {\it #2} {\bf #3} (#5) #4}}}
\newcommand{\doij}[5]{\href{http://dx.doi.org/#1}{{\cob {\it #2} {\bf #3} (#5) #4}}}

\newcommand{\procsin}[5]{in \emph{#1}, edited by #2 (#3, #4, #5)}
\newcommand{\tia}[1]{#1,}

\def\lp{\ell_{\rm Pl}}

\def\rmd{d}
\def\rmi{i}
\newcommand{\sgn}{\operatorname{sgn}} 

\begin{document}

\title{Multifractional spacetimes from the Standard Model to cosmology}

\author{Gianluca Calcagni}

\address{Instituto de Estructura de la Materia, CSIC, Serrano 121, 28006 Madrid, Spain\\ \email{calcagni@iem.cfmac.csic.es}}

\maketitle

\begin{history}
Submitted: February 5th, 2017

References update: September 22nd, 2017
\end{history}

\begin{abstract}
We review recent theoretical progress and observational constraints on multifractional spacetimes, geometries that change with the probed scale. On the theoretical side, the basic structure of the Standard Model and of the gravitational action is discussed. On the experimental side, we recall the bounds on the scales of the geometry coming from particle physics, astrophysics, and the cosmic microwave background.
\end{abstract}


\medskip

\authorfont{Mathematics Subject Classification 2010: 81Txx, 83D05}

%%%%%%%%%%%%%%%%%%%%%%%%%%%%%%%%%%%%%%%%%%%%%%%%%%%%%%%%%%%%%%%%%%%%%%%
%%%%%%%%%%%%%%%%%%%%%%%%%%%%%%%%%%%%%%%%%%%%%%%%%%%%%%%%%%%%%%%%%%%%%%%

\section{A landscape in quantum gravity}

This paper describes an example of the interaction between \emph{Geometry and Physics} (the topic of this XXV International Fall Workshop), both at the level of the mathematical structure and at the level of physical predictions. Multifractional theories are a clear case illustrating not only the restrictions one has to respect when modifying the smooth geometry of general relativity or quantum field theory, but also the conceptual and phenomenological possibilities one opens up when attempting to implement such modifications rigorously. But, first of all, why should one bother to change well-established paradigms?

A first motivation comes from theories of quantum gravity, all of which display a phenomenon called dimensional flow: the change of behaviour of correlation functions at different scales, which results in a spacetime with a scale-dependent dimension (Hausdorff dimension $\dh$ and/or spectral dimension $\ds$ and walk dimension $\dw$). This dimension usually runs to a value $<4$ in the ultraviolet (UV). This feature is universal in quantum gravity \cite{tH93,Car09} and has been studied extensively in specific frameworks such as perturbative effective quantum gravity, asymptotic safety, causal dynamical triangulations, Ho\v{r}ava--Lifshitz gravity, noncommutative spacetimes, nonlocal gravity, and the class of discrete combinatorial approaches made of loop quantum gravity, spin foams, and group field theory. A list of references and concrete examples can be found in \cite{revmu,Car17}. One question is whether dimensional flow has something to do with the UV finiteness (proven or putative) of these theories. For instance, can dimensional flow improve the perturbative renormalizability of a quantum field theory?

A second question, which was tentatively explored about 30 years ago, is whether dimensional flow is a mathematical property of these theories or if, on the contrary, can leave an observable imprint. Very preliminary results in the 1980s \cite{ScM,ZS,MuS} and more recent \cite{CO} showed that observations can constrain the value of the dimension of spacetime, but did not go beyond toy models. Placing phenomenological bounds on the dimension is all right, but it is desirable to have a theory justifying these explorations. First attempts in this direction tried to make sense of fractal spacetimes \cite{Sti77,Svo87,Ey89a,Ey89b}. A fractal spacetime is a spacetime whose dimension is noninteger; when scales are introduced in the geometry, we talk about multifractal spacetimes, which embody a special case of dimensional flow. These constructions were too technical to lead to manageable physical models and did not receive much attention. A more user-friendly revival of multifractal spacetimes \cite{fra1,fra2} soon met with other problems, such as the difficulty to define a self-adjoint Laplace--Beltrami operator \cite{revmu}.

Multifractional theories aim to give an answer to both questions. Concerning the first, by now it has become clear that dimensional flow does not guarantee UV finiteness or renormalizability, unless it is properly implemented. Much of the present and future research on this new paradigm is focused on the renormalizability properties of some of the multifractional theories, which are thus regarded as a stand-alone proposal independent of other quantum gravities. This study also helped to clarify many features of known quantum gravities thanks to a novel implementation of the techniques of fractal geometry and anomalous transport theory. Regarding the second question, the development of particle-physics and cosmological multifractional models gave rise to a battery of constraints on the scales at which dimensional-flow effects should become important. In turn, the feedback of these constraints on the parameter space of the theories helped to reformulate or clarify many of their aspects. This short review, written in a pedagogical way and without pretenses of being either formal or completely self-contained, is about these advances.

%%%%%%%%%%%%%%%%%%%%%%%%%%%%%%%%%%%%%%%%%%%%%%%%%%%%%%%%%%%%%%%%%%%%%%%
%%%%%%%%%%%%%%%%%%%%%%%%%%%%%%%%%%%%%%%%%%%%%%%%%%%%%%%%%%%%%%%%%%%%%%%

\section{ABC of multifractional spacetimes}

\begin{figure}
\centering
\includegraphics[width=8cm]{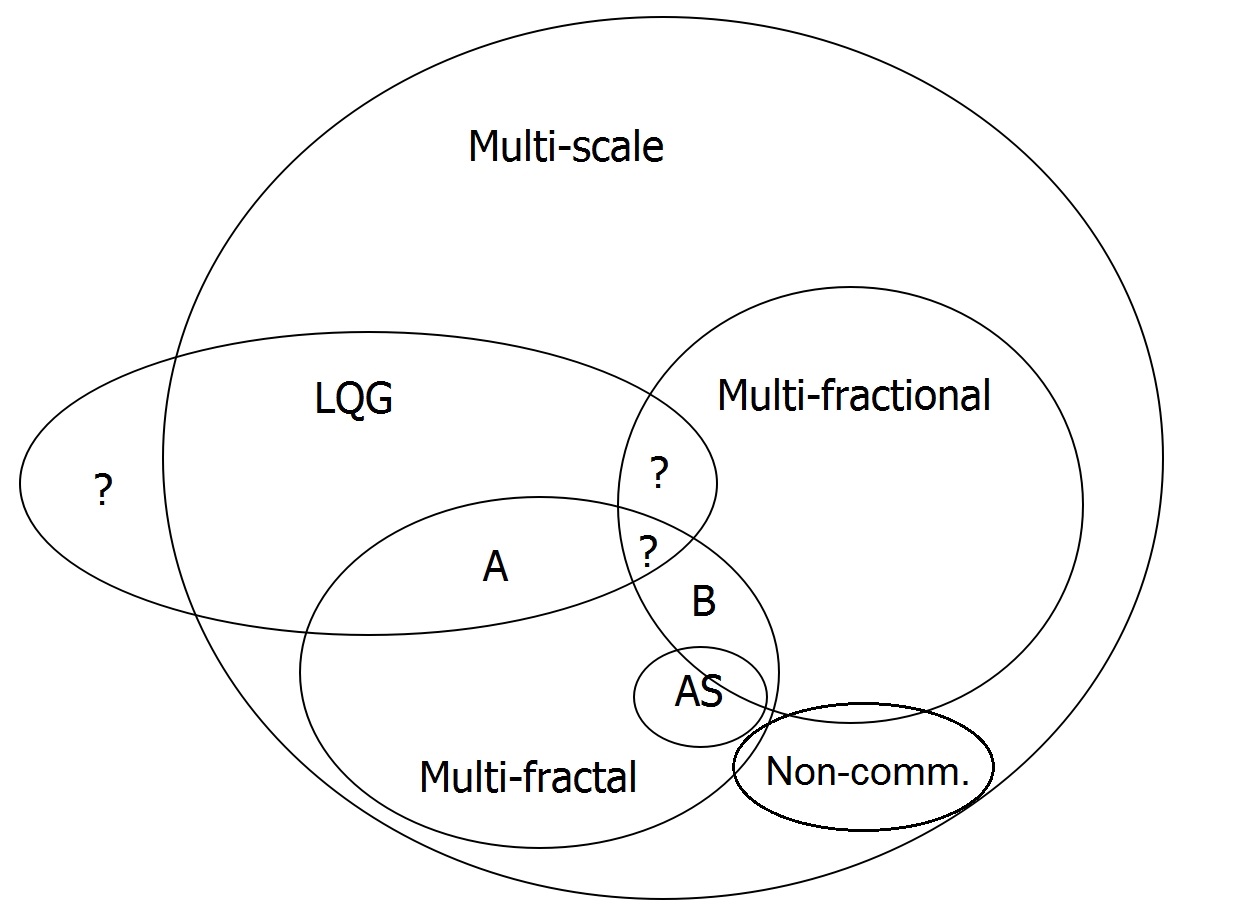}
\caption{The landscape of multiscale theories with anomalous geometry. Some theories predict geometries that are only multiscale, others that can be also multifractal, while in other theories (such as loop quantum gravity, spin foams, and group field theory) there also exist states of geometry that do not admit a well-defined spacetime dimension.}
\label{fig1}
\end{figure}

First, let us recall some definitions which have been recently put up to dissipate some confusion between multiscale, multifractal, and multifractional spacetimes \cite{revmu}. A scale-dependent spacetime geometry with realistic physical properties arises if the following conditions are met:
\begin{itemize}
\item[A.] Dimensional flow occurs, which happens if [A1] at least two of the dimensions $\dh$, $\ds$, and $\dw$ vary, [A2] the flow is continuous from the infrared (IR) to a UV cutoff, and [A3] the flow occurs locally.
\item[B.] An integer dimension observed at a finite number of points (e.g., at the UV and IR asymptotes of the flow).
\end{itemize}
These spacetimes are called \emph{multiscale}. Furthermore:
\begin{itemize}
\item[C.] If $\dw=2\dh/\ds$ and $\ds\leq\dh$ at all scales, we have a \emph{weakly multifractal spacetime}.
\item[D.] A geometry is a \emph{strongly multifractal spacetime} if, in addition of satisfying A--C, it is nowhere differentiable.
\end{itemize}
A simplified version of the landscape of multiscale and multifractal theories is shown in Fig.\ \ref{fig1}.

Multifractional theories describe multiscale (in some case, multifractal) spacetimes with certain symmetries. In a nutshell, one replaces the standard $D$-dimensional Lebesgue measure $\rmd^Dx$ with a nontrivial measure profile $\rmd^Dq(x)=\prod_{\mu=0}^{D-1}\rmd q^\mu(x^\mu)$. This measure is factorized in the coordinates $x^\mu$ for a purely technical reason: it allows us to build a self-adjoint Laplace--Beltrami operator whose eigenfunctions are the ``plane waves'' of a well-defined Fourier transform.

The action for some generic fields $\phi_f$ is
\be\label{ac}
\int\rmd^D x\,\cL[\p_x,\phi_f]\to \int\rmd^D{q(x)}\,\cL[{\cD_x},\phi_f]\,,
\ee
where $\cD_x$ is an integro-differential operator replacing first-order derivatives. The measure profiles $q^\mu(x^\mu)$, called geometric coordinates, are specific parametric distributions determined by a theorem \cite{first} stating that, \emph{if the spacetime Hausdorff dimension $\dh$ changes with the scale and if it reaches the IR value $D$ as a flat asymptote, then the most generic factorizable measure is a series dependent on an infinite hierarchy of scales; truncating the series to the largest scale $\ell_*^\mu$, one has}
\be
q^\mu(x^\mu) = \left[x^\mu+\frac{\ell_*^\mu\sgn(x^\mu)}{\a_\mu}\left|\frac{x^\mu}{\ell_*^\mu}\right|^{\a_\mu}\right]F_\om(x)\,,\label{q}
\ee
where $0<\a_\mu<1$ and $F_\om$ is a modulation factor characterized by logarithmic oscillations in the coordinates. Surprisingly, exactly the same profile is obtained when approximating integrals on deterministic multifractals with a continuous measure \cite{RLWQ,NLM,frc2}. Thus, if there is dimensional flow in $\dh$ and if this flow is slow in the IR (as it always is), then the measure is multiscale and, if we further assume factorizability, its \emph{binomial} approximation is \Eq{q}. From now on and to keep the presentation simple, we concentrate on the spatially \emph{isotropic} case where $\a_i=\a$, $\ell_*^0=t_*$, and $\ell_*^i=\ell_*$.

Let us see with a few examples some basic properties of multifractional geometry. $\bullet$ The simplest case, unphysical, is that of a one-dimensional fractional measure $q(x)\propto \sgn(x)|x/\ell_*|^\a$, giving $\rmd q(x)\propto \rmd x\,|x^\mu/\ell_*|^{\a-1}$. This geometry corresponds to a random fractal with dimension $\a$. By the scaling property $q(\la x)=\la^{{\a}} q(x)$ with arbitrary $\la$ (hence the adjective ``random'', as opposed to ``deterministic'' where $\la$ is fixed), one gets the Hausdorff dimension $\dh=\a$. $\bullet$ The $D$-dimensional isotropic case is straightforward (thanks to factorizability) and it gives $\dh=\a_0+(D-1)\a$, either by a self-similarity theorem or via the operational definition of $\dh$ as the scaling of the volume $\cV^{(D)}$ of a ball or of a hypercube of linear size $\ell$. $\bullet$ A third example is a Euclidean $D$-dimensional binomial measure with $\a_0=\a$ and without log oscillations ($F_\om=1$). In this case, the volume of the hypercube is made of two terms (hence the name ``binomial''), $\cV^{(D)}(\ell)= \ell_*^{D}[\Om_{D,1}({\ell}/{\ell_*})^{D}+\Om_{D,\a}({\ell}/{\ell_*})^{D\a}]$. For a small hypercube with ${\ell\ll\ell_*}$, the volume scales anomalously as $\cV^{(D)}\sim \tilde \ell^{D\a}$, where $\tilde \ell=\ell\ell_*^{-1+1/\a}$, while for $\ell\gg\ell_*$ we recover the usual scaling $\cV^{(D)}\sim \ell^{D\a}$. This geometry corresponds to a random multifractal with dimension varying from $D\a$ in the UV to $D$ in the IR. $\bullet$ Inclusion of log oscillations makes life more interesting. Let $D=1$. According to the flow-equation theorem \cite{revmu,first}, the measure at leading order is of the form [up to $\sgn(x)$ factors]
\be
q_{\a,{\om}}(x)= x+c_+ |x|^{\a+\rmi{\om}}+c_- |x|^{\a-\rmi{\om}},
\ee
where an argument from fractal geometry imposes $\om$ to take discretized values,
\be\label{ome}
\om=\om_N:=\frac{2\pi\a}{\ln N}\,,\qquad N=2,3,\dots\,.
\ee
Summing over $\a$ and $\om$ and imposing the measure to be real, one gets the full $D$-dimensional expression
\be
\rmd^D q(x) = \prod_\mu \left[\sum_n g_n\sum_\om\rmd q_{\a_n,\om}(x^\mu)\right]\,,
\ee
where
\be\label{qaom}
q_{\a,\om}(x) = \frac{x^\a}{\Gamma(\a+1)}\left[1+A\cos\left(\om\ln\frac{|x|}{{\ell_\infty}}\right)+B\sin\left(\om\ln\frac{|x|}{{\ell_\infty}}\right)\right]
\ee
and $A,B\in\mathbb{R}$ are amplitudes.\footnote{The sum in Eq.\ \Eq{qaom} should also extend to $A$ and $B$ in the most general case.} The quantity in square brackets was called $F_\om$ in Eq.~\Eq{q} and the scale $\ell_\infty$ can be identified with the Planck scale $\lp$ \cite{revmu}. This measure represents deterministic (in particular, self-similar) multifractals. The oscillatory part of $q(x)$ is {log-periodic} under the transformation $\ln({|x|}/{\ell_\infty})\to \ln({|x|}/{\ell_\infty})+{2\pi n}/{\om}$, $n=0,1,2,\dots$, implying the discrete scale invariance
\be
x\,\to\, \la_\om^n x\quad\Rightarrow\quad F_\om(\la_\om^n x)=F_\om(x)\,,\qquad \la_\om=\exp(-2\pi/\om)\,,\qquad n=0,1,2,\dots\,.
\ee
Discrete scale invariance appears not only in deterministic fractals, but also in chaotic systems \cite{Sor98}. 

The emergent picture is that of a fundamentally discrete geometry in the UV (measure with log oscillations) that is gradually coarse grained to a smooth multiscale geometry at mesoscopic scales (measure with $\langle F_\om\rangle=1$), which becomes standard spacetime in the IR (ordinary measure $q\simeq x$).

Due to factorizability and the scale dependence, the measure breaks all Poincaré symmetries. The derivative operator $\cD$ in the action \Eq{ac} is fixed by the symmetries imposed on the Lagrangian. It turns out that there are only three consistent possibilities:
\begin{enumerate}
\item Weighted derivatives: {$\cD_x=(\p_x q)^{-1/2}\p_x[(\p_x q)^{1/2}\,\cdot\,]$}. In this case, spacetime is not a multifractal and the spectral dimension is constant, $\ds=D$.
\item $q$-derivatives: {$\cD_x=\p_q=(\p_x q)^{-1}\p_x$}. In this case, spacetime is weakly multifractal.
\item Fractional derivatives: $\cD_x\sim\p^\a_x$. In this case, spacetime is strongly multifractal.
\end{enumerate}
In all cases, dimensional flow is implemented via a change of the integro-differential structure. The heavier the change, the more ``irregular'' the geometry.

The detailed study of these theories led to the following results:
\begin{itemize}
\item When regarded as effective models, multifractional theories (or, more generally, the multiscale formalism developed therein) capture the effective dynamics and the dimensional flow of several quantum gravities, noncommutative spacetimes, and varying-speed-of-light models.
\item When regarded as stand-alone models, the multifractional theories with $q$- and fractional derivatives can have some interesting renormalization properties (in particular, with relevance for quantum gravity) that avoid usual power counting \cite{revmu}. This result is very recent and preliminary. On the other hand, the theory with weighted derivatives does not have improved renormalizability.
\item Multifractional theories predict multiscale and discrete-geometry effects in virtually all sectors of physics, including cosmology, particle physics, and so on, all associated with easily falsifiable (much more easily than other quantum gravities) phenomenology.
\end{itemize}

Typically, when multifractional theories are presented at seminars, many questions arise in the audience regarding the fine print. These questions are collected in \cite{revmu} and answered there in detail. Here we limit our attention to a few basic but important aspects. One is about the possibility to regard the mapping $x\to q(x)$ as a coordinate change that trivializes the theory with $q$-derivatives (the most studied case). The answer is in the negative due to the fact that the presence of a scale hierarchy implies the existence of a preferred frame (called fractional frame and spanned by the coordinates $x^\mu$) where physical predictions are computed. In this frame, clocks and rods are scale-independent objects with which we measure reality and their scale independence is a reflection of our own scale specificity. The frame spanned by the composite coordinates $q$ (called integer frame) would correspond to scale-dependent (``adaptive'') apparatus. The existence of a preferred frame can be understood by the way multiscaling affects relational measurements. Consider, for example, the velocities $V_x=\Delta x/\Delta t$ (in the fractional frame) and $V_q=\Delta q(x)/\Delta q(t)$ (in the integer frame). Choosing measurement units is a convention and measuring a size or a velocity by an observer in a multifractal world is not different from measuring them in a normal nonanomalous world. However, measurements of dimensionless observables do discriminate between standard and multiscale geometries, and the ratios $V_x(O_1)/V_x(O_2)$ and $V_q(O_1)/V_q(O_2)$ are typically different for two observers in hierarchical order, ${\rm scale}(O_1)\ll {\rm scale}(O_2)$.

A self-interacting scalar field in two-dimensional flat space may further help to clarify the frame choice. In the theory with $q$-derivatives, we have
\ba
S_\phi&=&\int\rmd^2 q\,\left\{\frac12[\p_{q^0(t)}\phi]^2-\frac12[\p_{q^1(x)}\phi]^2-\sum_n\la_n\phi^n\right\}\label{f1}\\
&=&\int\rmd^2 x\,\left\{\vphantom{\sum_n}\frac{v_1(x)}{2v_0(t)}\dot\phi^2-\frac{v_0(t)}{2v_1(x)}(\p_x\phi)^2-\sum_n[v_0(t)v_1(x)\la_n]\phi^n\right\},\label{f2}
\ea
where $v_1=\p_{x^1} q^1$ and $v_0=\p_t q^0$. The integer picture [Eq.\ \Eq{f1} with coordinate dependence of $q^0$ and $q^1$ removed] is used to perform intermediate calculations. In this frame, any ``time'' or ``spatial'' interval or ``energy'' are measured with $q$-clocks, $q$-rods, or $q$-detectors, and these fictitious observables must be reconverted to the fractional picture to get physical observables.

The theory with weighted derivatives has a similar frame choice but, in this case, the relation between the fractional and the integer frame is a field redefinition rather than a coordinate mapping:
\ba
S_\phi&=&\int\rmd^2x\,v\left[\frac12(\cD_t\phi)^2-\frac12(\cD_{x}\phi)^2-\sum_n\la_n\phi^n\right]\nonumber\\
&=&\int\rmd^2 x\,\left[\frac12(\p_t\vp)^2-\frac12(\p_{x}\vp)^2-\sum_n v^{1-\frac{n}{2}}\la_n\vp^n\right],\quad  \vp=\sqrt{v}\phi,
\ea
where $v=v_0v_1$. Again, one can work intermediate steps in the integer picture but, now, that frame is not identical to a standard theory due to self-interactions. Even in the absence of self-interactions, it is impossible to remove multiscaling effects in physical observables, due to the fundamental difference between Lagrangian couplings and physical couplings (charges of Noether currents) in particle physics \cite{frc13}. In the presence of gravity, the integer picture becomes even more complicated and it clearly deviates from standard.

The theory with fractional derivatives is highly nontrivial and probably does not admit an integer-frame description. We will not discuss this case here.

Other questions about the structure of the measure and the corresponding physical interpretation [Does $q(x)\to q(x-\bar x)$ lead to a different theory? To a different physics? What is the interpretation of the special point $\bar x$?] are related to the frame choice and have been settled recently by the flow-equation theorem \cite{first}, which, as said above, severely limits the form of $q(x)$. The reader can find a full discussion in \cite{revmu}.

We need one last piece of information before moving to physical models. Momentum space reflects the scale dependence of geometry in a complementary way with respect to position space. In the theory with $q$-derivatives, the existence of an invertible Fourier transform fixes the momentum measure $\rmd^Dp(k)=\rmd p^0(E)\,\rmd p^1(k^1)\dots$ by the relation (index $\mu$ omitted)
\be
p(k)=\frac{1}{q(1/k)}=\frac{k}{1+\frac{1}{\a}\left|\frac{E_*}{k}\right|^{\a-1} F_\om(k)}\,,
\ee
where spatial momentum scales have been all identified with some $E_*$ for the sake of illustration. The description in momentum coordinates involve one or more fundamental {energy scales} and there is no reference to any special point in space.

%%%%%%%%%%%%%%%%%%%%%%%%%%%%%%%%%%%%%%%%%%%%%%%%%%%%%%%%%%%%%%%%%%%%%%%
%%%%%%%%%%%%%%%%%%%%%%%%%%%%%%%%%%%%%%%%%%%%%%%%%%%%%%%%%%%%%%%%%%%%%%%

\section{Standard Model, gravity, and cosmology}

In this section, we recall some facts about electroweak, strong, and gravitational interactions.

The Standard Model action in the theory with weighted derivatives is the usual one with the replacements
\be
\p_\mu\to\cD_\mu\,,\qquad {\rm (coupling)}\to \sqrt{v}\,\,{\rm (coupling)}\,.
\ee
The full action can be found in \cite{revmu,frc13}. Time and lengths in the integer picture are the usual ones. Studying quantum electrodynamics (QED), it turns out that the fine-structure constant acquires, via the electric charge, a scale dependence in the time direction:
\be
\a_\textsc{qed}(t)=\frac{\tilde\a_\textsc{qed}}{v_0(t)},
\ee
where $\tilde\a_\textsc{qed}$ is the fine-structure constant (really constant) in the integer picture. This time dependence in a coupling of Nature has been constrained by experiments, including on the allowed variation of $\a_\textsc{qed}$ and the Lamb shift effect \cite{frc13}.

The Standard Model with $q$-derivatives is obtained from the usual action with the replacement $\p_\mu\to\p_{q^\mu}$ everywhere. The resulting action \cite{frc13} is invariant under $q$-Poincar\'e transformations ${q'}^\mu(x^\mu)=\Lambda_\nu^{\ \mu}q^\nu(x^\nu)+a^\mu$ and also under CPT. Times and lengths in the integer picture are {composite} and give rise to scale-dependent observables in the fractional picture. An example is muon lifetime $\tau_{\rm mu}$, which is obtained from the usual expression $\tau_0$ by inverting
\be
q^0(\tau_{\rm mu})=\frac{1}{\Gamma}=\tau_0,\qquad \Gamma = \frac{G^2_{\rm F} m_{\rm mu}^5}{192\pi^3}+\cdots\,,
\ee
where $G_{\rm F}$ is Fermi constant and $m_{\rm mu}$ is the muon mass.

The gravitational action of the theories with weighted and $q$-derivatives was presented in \cite{frc11}. The first case is interesting because it strongly resembles a Weyl-integrable spacetime where the metric is not covariantly conserved. Here we will only discuss the case with $q$-derivatives. Defining the multiscale Levi-Civita connection and Riemann tensor
\ba
&&{}^q\G^\rho_{\mu\nu} := \tfrac12 g^{\rho\s}\left(\frac{1}{v_\mu}\p_{\mu} g_{\nu\s}+\frac{1}{v_\nu}\p_{\nu} g_{\mu\s}-\frac{1}{v_\s}\p_\s g_{\mu\nu}\right)\,,\label{leciq}\\
&&{}^q R^\rho_{~\mu\s\nu}:= \frac{1}{v_\s}\p_\s {}^q\G^\rho_{\mu\nu}-\frac{1}{v_\nu}\p_\nu {}^q\G^\rho_{\mu\s}+{}^q\G^\tau_{\mu\nu}\,{}^q\G^\rho_{\s\tau}-{}^q\G^\tau_{\mu\s}\,{}^q\G^\rho_{\nu\tau}\,,
\ea
one obtains the deceptively simple action
\be\label{Sgq}
S =\frac{1}{2\kappa^2}\int\rmd^Dx\,v\,\sqrt{-g}\,({}^q R-2\Lambda)+S_{\rm m}\,,
\ee
where we added a cosmological constant and $S_{\rm m}$ is the matter action. The metric $g_{\mu\nu}$ is completely independent of the measure structure, although they are related to each other through the dynamical equations
\be\label{eeq}
{}^qR_{\mu\nu}-\frac12 g_{\mu\nu} ({}^q R-2\Lambda)=\kappa^2\, {}^qT_{\mu\nu}\,,
\ee
where ${}^qT_{\mu\nu}$ is the matter energy-momentum tensor. The presence of a fixed nondynamical structure independent of the metric structure explicitly breaks the equivalence between covariance and diffeomorphism invariance typical of general relativity \cite{Giu06}. Diffeomorphism invariance itself is preserved in this theory albeit with a new twist. Also, a local inertial frame centered on the observer is locally isomorphic to multifractional Minkowski spacetime and each and every local inertial frame has its own anomalous distribution $q(x)$.

The cosmology of the $q$-theory is simple to work out. For example, the first Friedmann equation for a homogeneous and isotropic background is
\be
\frac{H^2}{v^2}=\frac{\kappa^2}{3}\,\rho+\frac{\Lambda}{3}-\frac{\textsc{k}}{a^2}\,,\label{fri}
\ee
where $\k^2=8\pi G$ is Newton constant, $a$ is the scale factor, $H=\dot a/a$ is the Hubble parameter, and $\textsc{k}$ is the curvature of spatial slices. A power-law solution $a(t)=[q^0(t)]^p$ reveals one of the most characteristic features of multifractional cosmology: the universe does not expand monotonically but undergoes an infinite sequence of cycles of contractions and expansions with increasing frequency towards the past. At late times $t\gg t_*$, the log-oscillating modulation and multiscale effects disappear and standard expansion is recovered.

%Another aspect to highlight is that the ordinary {slow-roll} approximation is {unnecessary} in order to get an accelerated expansion. The intuitive reason is that the measure structure affects the scale factor via the dynamics in such a way that, in certain regimes, the Hubble horizon $H^{-1}$ can look like shrinking for an observer.
%In a power-law cosmology, for ordinary matter with equation of state $w>-1/3$, the continuity equation $\dot\rho+3H(1+w)\rho=0$ and the Friedmann equation \Eq{fri} imply that $p<1$. On the other hand, one has $\ddot a=p (q^0)^{p-2}[q^0\ddot q^0+(p-1)(\dot q^0)^2]$, which is negative in standard cosmology [$q^0=t$, $\ddot a=p(p-1)t^{p-2}$]. However, the nontrivial part of $q^0$ contributes with a positive term to $\ddot a$ which
%q\ddot q+(p-1)\dot q^2 = (p-1)+(2p+\a-3)|t_*/t|^{1-\a}+(p-1/\a)|t_*/t|^{2(1-\a)}
%for $t_*/t\approx 1$, (p-1)+(2p+\a-3)+(p-1/\a)=4(p-1)+\a-1/\a<0
%for $t_*/t\gg 1$  (p-1/\a)|t_*/t|^{2(1-\a)}<0
 %Technically, the left-hand side of \Eq{fri} is not constant even when $H$ is.

Inflationary scalar and tensor spectra can be worked out and their general behaviour at or above scales $E_*$ is \cite{frc11,frc14}
\be
P_{\rm s,t} \sim k^{\a n} (1+\text{log oscillations})\,,\label{Pst}
\ee
where $n=n_{\rm s}-1,n_{\rm t}$ is the spectral index in, respectively, the scalar and tensor sector. Two major consequences one can draw from \Eq{Pst} are:
\begin{enumerate}
\item Since the effective spectral index at large scales is $\a n$ and it can be close to zero even when $n$ is not (because $0<\a<1$), one can get almost scale invariance even when the slow-roll approximation is softened.
\item A {log-oscillating pattern} in the spectrum, deforming the usual sequence of peaks and troughs, arises. This is a direct imprint in the sky of the discrete spacetime geometry at scales $\sim \ell_\infty$. Its visible effect is not ``holes'' in the fabric of spacetime but a logarithmic modulation of the power spectrum of primordial fluctuations.
\end{enumerate}

%%%%%%%%%%%%%%%%%%%%%%%%%%%%%%%%%%%%%%%%%%%%%%%%%%%%%%%%%%%%%%%%%%%%%%%
%%%%%%%%%%%%%%%%%%%%%%%%%%%%%%%%%%%%%%%%%%%%%%%%%%%%%%%%%%%%%%%%%%%%%%%

\section{Constraints on scales}

Having got a bird's eye view on multifractional theories, we present some observational constraints according to the physics sector where they come from. In all cases, the idea is to use the experimental error associated with an observable as an upper bound on the magnitude of multiscale effects, which leads to an upper bound on the time and length scales $t_*$ and $\ell_*$ and to a lower bound on the energy scale $E_*$. Since smaller length scales would produce subdominant effects, it is not necessary to consider measures more complicated than the binomial profile \Eq{q}. The scale $\ell_\infty=\lp$ appearing in log oscillations is a different matter, since it induces a long-range modulation of multiscale effects even when it is much smaller than $\ell_*$.

We divide the type of bounds into absolute constraints and constraints for a given $\a_0$ or $\a$. Absolute constraints are usually obtained for very small $\a_0,\a$ and are the most conservative. Bounds for a given $\a_0$ or $\a$ are far more restrictive if these parameters are not too close to zero.

The type of measurement involved will be divided into particle-physics, astrophysical, and cosmological. Details are given in \cite{revmu}.

%%%%%%%%%%%%%%%%%%%%%%%%%%%%%%%%%%%%%%%%%%%%%%%%%%%%%%%%%%%%%%%%%%%%%%%

\subsection{Theory with weighted derivatives}

For this theory, the only extant absolute bounds in particle physics come from quantum electrodynamics (Tab.\ \ref{tab1}). Astrophysical processes cannot constraint this theory because it predicts an ordinary dispersion relation. The only astrophysical bound is from the variation of the fine-structure constant in quasar observations \cite{frc8}, but it is insignificant ($E_*>10^{-28}\,{\rm eV}$). 
\begin{table}[ht]
\begin{center}
\tbl{\label{tab1}Particle-physics absolute bounds on the scales of the multifractional theory $T_v$ with weighted derivatives.}
{\begin{tabular}{l|ccc}\hline\hline
$T_v$ ($\a_0,\a\ll 1/2$)  							          & $t_*$ (s)        & $\ell_*$ (m) & $E_*$ (GeV) \\\hline
Lamb shift       																  & ${<10^{-23}}$& $<10^{-14}$  & $>10^{-2}$      \\
Measurements of $\a_\textsc{qed}$ 							  & ${<10^{-26}}$& $<10^{-18}$  & $>10^{1}$       \\\hline\hline
\end{tabular}}
\end{center}
\end{table}

To get specific bounds, one must pick a value of the exponents $\a_0$ and $\a$ in the measure. The value $1/2$ lying right in the middle of the allowed interval $(0,1)$ is a typical choice. The electrodynamics constraints for $\a_0=1/2=\a$ are reported in Tab.\ \ref{tab2}. On the astrophysics side, again, the bound from quasars is too weak \cite{frc8}. There is also a bound coming from the cosmic microwave background (CMB) black-body spectrum \cite{frc14}, but it is much weaker than those in the table. 
\begin{table}[ht]
\begin{center}
\tbl{\label{tab2}Particle-physics bounds on the scales of the multifractional theory $T_v$ with weighted derivatives for $\a_0=1/2=\a$.}
{\begin{tabular}{l|ccc}\hline\hline
$T_v$	($\a_0=1/2=\a$)							                & $t_*$ (s)     		& $\ell_*$ (m) & $E_*$ (GeV)\\\hline
Lamb shift       																  & ${<10^{-29}}$ & $<10^{-20}$  & $>10^4$         \\
Measurements of $\a_\textsc{qed}$ 								& ${<10^{-36}}$ & $<10^{-28}$  & $>10^{11}$    	 \\\hline\hline
\end{tabular}}
\end{center}
\end{table}
Apart from this, we have no other cosmological bounds (absolute or specific) simply because they have not been calculated yet.

The strongest bound comes from measurements of the fine-structure constant \cite{frc13}. The latter is measured with accuracy ${\delta\a_\textsc{qed}}/{\a_\textsc{qed}}\sim 10^{-10}$. Since $\a_\textsc{qed}(t)={\tilde\a_\textsc{qed}}/{v_0(t)}$ in this theory, the difference between the usual and the multifractional fine-structure constant is $\Delta\a_\textsc{qed}=\a_\textsc{qed}(t)|t_*/t|^{1-\a_0}$, where we ignored log oscillations. Imposing $\Delta\a_\textsc{qed}<\delta\a_\textsc{qed}$ and taking times $t\sim t_\textsc{qed}=10^{-16}\,{\rm s}$ of order of the typical QED interaction, one gets
\be
t_*< 10^{-16-10/(1-\a_0)}\,{\rm s}\,.
\ee
This is the source of time bounds in the last line of the tables. Length and energy bounds are obtained by a unit conversion. 

To summarize, the best result so far for this theory is that the fundamental energy scale is bounded from below by
\be\label{bo1}
E_*>10\,\text{GeV}\,,\qquad E_*^{(\a_0=1/2)}>10^{11}\,\text{GeV}\,.
\ee
The most conservative bound is rather weak but, when $\a_0$ increases, it rapidly increases to interesting values around the grand-unification scale and beyond.

%%%%%%%%%%%%%%%%%%%%%%%%%%%%%%%%%%%%%%%%%%%%%%%%%%%%%%%%%%%%%%%%%%%%%%%

\subsection{Theory with \texorpdfstring{$q$}{}-derivatives}

The particle-physics constraints for this theory are richer because also the weak sector is affected by anomalous geometry. The absolute and $\a_0=1/2=\a$ bounds are given in Tab.\ \ref{tab3}. Contrary to the case with weighted derivatives, here the fine-structure constant coincides with the usual one and there is no such a strong bound as \Eq{bo1}.
\begin{table}[ht]
\begin{center}
\tbl{\label{tab3} Particle-physics absolute bounds (obtained for $\a_0,\a\ll 1$, upper part) and bounds for $\a_0=1/2=\a$ (lower part) on the scales of the multifractional theory $T_q$ with $q$-derivatives.}
{\begin{tabular}{l|ccc}\hline\hline
$T_q$	($\a_0,\a\ll 1/2$)						              & $t_*$ (s)         & $\ell_*$ (m) & $E_*$ (GeV) \\\hline
Muon lifetime    																  & ${<10^{-11}}$ & $<10^{-3}$   & $> 10^{-13}$\\
Lamb shift        																& $<10^{-21}$       & $<10^{-13}$  & ${>10^{-4}}$\\\hline\hline
$T_q$	($\a_0=1/2=\a$)							                & $t_*$ (s)     		& $\ell_*$ (m) & $E_*$ (GeV)\\\hline
muon lifetime    																  & ${<10^{-17}}$ & $<10^{-8}$   & $> 10^{-8}$ \\
Lamb shift        																& $<10^{-26}$       & $<10^{-18}$  & ${>10^1}$   \\\hline\hline
\end{tabular}}
\end{center}
\end{table}

Cosmological bounds are complementary to those from the Standard Model. The CMB black-body spectrum yields constraints of the scales $t_*,\ell_*,E_*$ of the same order of magnitude as the Lamb shift, while the CMB primordial scalar spectrum is not very sensitive to these scales \cite{frc14}. However, the log oscillations in \Eq{Pst} can easily disrupt the delicate balance in parameter space achieved by the standard power-law spectrum, which simply means that the wealth of constraints on such parameter space limit severely the amplitudes $A$ and $B$ in \Eq{qaom}. Fixing $N$ in \Eq{ome} and $\a=1/2$, one finds \cite{frc14}
\be\label{AB}
\begin{matrix}
N=2:\qquad & A<0.3,\,B<0.4\,,\\
N=3:\qquad & A<0.3,\,B<0.2\,,\\
N=4:\qquad & A<0.4,\,B<1.0\,.
\end{matrix}
\ee
When $\a$ is not marginalized, a likelihood analysis finds an interesting upper bound on it, $\a\lesssim 0.1-0.6$ for $N=2,3,4$, which leads to a curious \emph{upper bound} on the UV spatial Hausdorff dimension $\dh^{\rm \,space}\simeq 3\a$:
\be\label{dhbo}
N=2,3,4:\qquad \dh^{\rm \,space}\lesssim 0.3-1.9\qquad\text{(UV)}\,.
\ee
This counter-intuitive result states that, as soon as we allow for a nontrivial dimensional flow in $\dh$, the UV value cannot be arbitrarily close to the IR value 4. The bounds \Eq{AB}
(on the discrete UV structure of spacetime) and \Eq{dhbo} (on the UV spacetime dimension) are the first of this kind in quantum gravity.

However, very powerful constraints from astrophysics put this theory in jeopardy (Tab.\ \ref{tab4}). These bounds entail the measurement of the difference between the propagation speed of different particles (protons and photons in the Cherenkov radiation case) or of the same type of particles but with different energies (photons in the gamma-ray burst case). In general, for a massless particle the theory with $q$-derivatives predicts a modified dispersion relation of the form \cite{revmu}
\be
E^2\simeq k^2\left[1\pm O(1)\left(\frac{k}{E_*}\right)^{1-\a}\right],\qquad 0<1-\a<1,
\ee
which differs from traditional quantum-gravity-motivated and string-inspired dispersion relations $E^2 \simeq k^2 [1+b({k}/{M})^n]$ in the range of the exponent of the correction, between 0 and 1 in the multifractional case and equal to $n=1$ or $n=2$ \cite{ACEMN,ACAP,ArCa2} in all the other cases. Therefore, the correction to the usual dispersion relation $E^2=k^2$ is less suppressed in the multifractional case and the energy scale $E_*$ is better constrained by observations than the mass scale $M$.
\begin{table}[ht]
\begin{center}
\tbl{\label{tab4}Astrophysics absolute bounds (obtained for $\a_0,\a\ll 1$, upper part) and bounds for $\a_0=1/2=\a$ (lower part) on the scales of the multifractional theory $T_q$ with $q$-derivatives. GRB stands for gamma-ray burst.}
{\begin{tabular}{l|ccc}\hline\hline
$T_q$	($\a_0,\a\ll 1/2$)						              & $t_*$ (s)         & $\ell_*$ (m) & $E_*$ (GeV)   \\\hline
GRBs 																						  & $<10^{-39}$       & $<10^{-30}$  & ${>10^{14}}$\\
vacuum Cherenkov radiation 														  & $<10^{-57}$				& $<10^{-49}$  & ${>10^{33}}$\\\hline\hline
$T_q$	($\a_0=1/2=\a$)							                & $t_*$ (s)     		& $\ell_*$ (m) & $E_*$ (GeV)  \\\hline
GRBs																							& $<10^{-57}$       & $<10^{-48}$  & ${>10^{32}}$\\
vacuum Cherenkov radiation																& $<10^{-79}$     	& $<10^{-71}$  & ${>10^{55}}$\\\hline\hline
\end{tabular}}
\end{center}
\end{table}

The constraint on vacuum Cherenkov radiation emitted by ultra-high-energy cosmic rays has been found very recently \cite{revmu} and it completely rules out the theory.

%%%%%%%%%%%%%%%%%%%%%%%%%%%%%%%%%%%%%%%%%%%%%%%%%%%%%%%%%%%%%%%%%%%%%%%

\subsection{Theory with fractional derivatives}

The theory $T_{\g}$ with fractional derivatives is still under construction and it is too early to say anything robust about its phenomenology. However, one can get a preliminary idea by noting that the theory with $q$-derivatives can be regarded also as an approximation (denoted as $T_{\g=\a}\cong T_q$) of the theory with fractional derivatives in the so-called deterministic view \cite{revmu}. This subcase with fractional derivatives is, in principle, disfavored by the above results, unless some nontrivial effect not accounted for by the approximation $T_{\g=\a}\cong T_q$ enter the game; only the full construction of $T_\g$ will be able to settle the issue \cite{revmu}. On the other hand, there exists a second subtype of $T_\g$ where the fabric of spacetime is intrinsically stochastic. In this case, multiscale corrections are interpreted as stochastic fluctuations and, if these cancel out in average, all the strongest bounds (including from GRBs and vacuum Cherenkov radiation) would be avoided. This possibility has been explored recently \cite{CaRo2}.

\section*{Acknowledgments}

The author is supported by the I+D grants FIS2014-54800-C2-2-P and FIS2017-86497-C2-2-P of the Ministry of Science, Innovation and Universities.


\begin{thebibliography}{99}
\bibitem{tH93}  G.\ 't Hooft, \tia{Dimensional reduction in quantum gravity} \procsin{Salamfestschrift}{A.\ Ali, J.\ Ellis, and S.\ Randjbar-Daemi}{World Scientific}{Singapore}{1993} [\oarX{gr-qc/9310026}].
\bibitem{Car09} S.\ Carlip, \tia{Spontaneous dimensional reduction in short-distance quantum gravity?} \doinn{10.1063/1.3284402}{AIP Conf.\ Proc.}{1196}{72}{2009} [\arX{0909.3329}].
\bibitem{revmu} G.\ Calcagni, \tia{Multifractional theories: an unconventional review} \doij{10.1007/JHEP03(2017)138}{J.\ High Energy Phys.}{1703}{138}{2017} [\arX{1612.05632}]
\bibitem{Car17} S.\ Carlip, \tia{Dimension and dimensional reduction in quantum gravity} \doinn{10.1088/1361-6382/aa8535}{Class.\ Quantum Grav.}{34}{193001}{2017} [\arX{1705.05417}].
\bibitem{ScM}   A.\ Sch\"afer and B.\ M\"uller, \tia{Bounds for the fractal dimension of space} \doin{10.1088/0305-4470/19/18/034}{J.\ Phys.}{A}{19}{3891}{1986}.
\bibitem{ZS}    A.\ Zeilinger and K.\ Svozil, \tia{Measuring the dimension of space-time} \doinn{10.1103/PhysRevLett.54.2553}{Phys.\ Rev.\ Lett.}{54}{2553}{1985}.
\bibitem{MuS}   B.\ M\"uller and A.\ Sch\"afer, \tia{Improved bounds on the dimension of space-time} \doinn{10.1103/PhysRevLett.56.1215}{Phys.\ Rev.\ Lett.}{56}{1215}{1986}.
\bibitem{CO}    F.\ Caruso and V.\ Oguri, \tia{The cosmic microwave background spectrum and a determination of fractal space dimensionality} \doinn{10.1088/0004-637X/694/1/151}{Astrophys.\ J.}{694}{151}{2009} [\arX{0806.2675}].
\bibitem{Sti77} F.H.\ Stillinger, \tia{Axiomatic basis for spaces with noninteger dimension} \doinn{10.1063/1.523395}{J.\ Math.\ Phys.}{18}{1224}{1977}.
\bibitem{Svo87} K.\ Svozil, \tia{Quantum field theory on fractal space-time} \doin{10.1088/0305-4470/20/12/033}{J.\ Phys.}{A}{20}{3861}{1987}.
\bibitem{Ey89a} G.\ Eyink, \tia{Quantum field-theory models on fractal spacetime. I: Introduction and overview} \doinn{10.1007/BF01228344}{Commun.\ Math.\ Phys.}{125}{613}{1989}.
\bibitem{Ey89b} G.\ Eyink, \tia{Quantum field-theory models on fractal spacetime. II: Hierarchical propagators} \doinn{10.1007/BF02124332}{Commun.\ Math.\ Phys.}{126}{85}{1989}.
\bibitem{fra1}  G.\ Calcagni, \tia{Fractal universe and quantum gravity} \doinn{10.1103/PhysRevLett.104.251301}{Phys.\ Rev.\ Lett.}{104}{251301}{2010} [\arX{0912.3142}].
\bibitem{fra2}  G.\ Calcagni, \tia{Quantum field theory, gravity and cosmology in a fractal universe} \doij{10.1007/JHEP03(2010)120}{J.\ High Energy Phys.}{1003}{120}{2010} [\arX{1001.0571}].
\bibitem{first} G.\ Calcagni, \tia{Multiscale spacetimes from first principles} \doin{10.1103/PhysRevD.95.064057}{Phys.\ Rev.}{D}{95}{064057}{2017} [\arX{1609.02776}].
\bibitem{RLWQ}  F.-Y.\ Ren, J.-R.\ Liang, X.-T.\ Wang, and W.-Y.\ Qiu, \tia{Integrals and derivatives on net fractals} \doinn{10.1016/S0960-0779(02)00211-4}{Chaos Solitons Fractals}{16}{107}{2003}.
\bibitem{NLM}   R.R.\ Nigmatullin and A.\ Le M\'ehaut\'e, \tia{Is there geometrical/physical meaning of the fractional integral with complex exponent?} \doinn{10.1016/j.jnoncrysol.2005.05.035}{J.\ Non-Cryst.\ Solids}{351}{2888}{2005}.
\bibitem{frc2}  G.\ Calcagni, \tia{Geometry and field theory in multi-fractional spacetime} \doij{10.1007/JHEP01(2012)065}{J.\ High Energy Phys.}{1201}{065}{2012} [\arX{1107.5041}].
\bibitem{Sor98} D.\ Sornette, \tia{Discrete scale invariance and complex dimensions} \doinn{10.1016/S0370-1573(97)00076-8}{Phys.\ Rep.}{297}{239}{1998} [\oarX{cond-mat/9707012}].
\bibitem{frc13} G.\ Calcagni, G.\ Nardelli, and D.\ Rodr\'iguez-Fern\'andez, \tia{Standard Model in multiscale theories and observational constraints} \doin{10.1103/PhysRevD.94.045018}{Phys.\ Rev.}{D}{94}{045018}{2016} [\arX{1512.06858}].
\bibitem{frc11} G.\ Calcagni, \tia{Multi-scale gravity and cosmology} \doij{10.1088/1475-7516/2013/12/041}{J.\ Cosmol.\ Astropart. Phys.}{12}{041}{2013} [\arX{1307.6382}].
\bibitem{Giu06} D.\ Giulini, \tia{Some remarks on the notions of general covariance and background independence} \doinn{10.1007/978-3-540-71117-9_6}{Lect.\ Notes Phys.}{721}{105}{2007} [\oarX{gr-qc/0603087}].
\bibitem{frc14} G.\ Calcagni, S.\ Kuroyanagi, and S.\ Tsujikawa, \tia{Cosmic microwave background and inflation in multi-fractional spacetimes} \doij{10.1088/1475-7516/2016/08/039}{J.\ Cosmol.\ Astropart.\ Phys.}{1608}{039}{2016} [\arX{1606.08449}].
\bibitem{frc8}  G.\ Calcagni, J.\ Magueijo, and D.\ Rodr\'iguez-Fern\'andez, \tia{Varying electric charge in multiscale spacetimes} \doin{10.1103/PhysRevD.89.024021}{Phys.\ Rev.}{D}{89}{024021}{2014} [\arX{1305.3497}].
\bibitem{ACEMN} G.\ Amelino-Camelia, J.R.\ Ellis, N.E.\ Mavromatos, and D.V.\ Nanopoulos, \tia{Distance measurement and wave dispersion in a Liouville string approach to quantum gravity} \doin{10.1142/S0217751X97000566}{Int.\ J.\ Mod.\ Phys.}{A}{12}{607}{1997} [\oarX{hep-th/9605211}].
\bibitem{ACAP}  G.\ Amelino-Camelia, M.\ Arzano, and A.\ Procaccini, \tia{Severe constraints on loop-quantum-gravity energy-momentum dispersion relation from black-hole area-entropy law} \doin{10.1103/PhysRevD.70.107501}{Phys.\ Rev.}{D}{70}{107501}{2004} [\oarX{gr-qc/0405084}].
\bibitem{ArCa2} M.\ Arzano and G.\ Calcagni, \tia{What gravity waves are telling about quantum spacetime} \doin{10.1103/PhysRevD.93.124065}{Phys.\ Rev.}{D}{93}{124065}{2016} [\arX{1604.00541}].
\bibitem{CaRo2} G.\ Calcagni and M.\ Ronco, \tia{Dimensional flow and fuzziness in quantum gravity: emergence of stochastic spacetime} \doin{10.1016/j.nuclphysb.2017.07.016}{Nucl.\ Phys.}{B}{923}{144}{2017} [\arX{1706.02159}].
\end{thebibliography}
\end{document}